\documentclass[superscriptaddress,onecolumn]{revtex4}
\usepackage{graphicx,epsfig}
\usepackage{amsmath}
\usepackage {amssymb}
\usepackage{multirow}
\usepackage{xfrac}
\usepackage{ulem}
\usepackage{float}
\usepackage{color}

\newcommand{\be}{\begin{eqnarray}}
\newcommand{\ee}{\end{eqnarray}}
\newcommand{\bea}{\begin{eqnarray}}

\newcommand{\eea}{\end{eqnarray}}

\def\a{\alpha}

\def\beq{\begin{equation}}
\def\eeq{\end{equation}}
\def\beqn{\begin{eqnarray}}
\def\eeqn{\end{eqnarray}}
\def\ba{\begin{eqnarray}}
\def\ea{\end{eqnarray}}

\def\xprim2bar{\overline{x}^{\prime\prime}}

\def\beq{\begin{equation}}
\def\eeq{\end{equation}}

\def\p{\partial}
\def\pr{\prime}
\def\pp{\prime\prime}
\def\pp{{\prime\prime}}

\newcommand{\f}{\frac}

\def\A5{(A_5)_{\rm lat}}

\makeindex

\begin{document}

\title{Thermal Stability and QNMs of a Hairy Black Hole in the Presence of a Monopole Field}

\author{George Koutsoumbas}
	\email{kutsubas@central.ntua.gr}
	\affiliation{Physics Division, School of Applied Mathematical and Physical Sciences, National Technical University of Athens, 15780 Zografou Campus,
    Athens, Greece.}

\author{Andri Machattou}
	\email{andrimachattou@hotmail.com}
	\affiliation{Physics Division, School of Applied Mathematical and Physical Sciences, National Technical University of Athens, 15780 Zografou Campus,
    Athens, Greece.}
	
	\author{Eleftherios Papantonopoulos}
         \email{lpapa@central.ntua.gr}
	\affiliation{Physics Division, School of Applied Mathematical and Physical Sciences, National Technical University of Athens, 15780 Zografou Campus,
    Athens, Greece.}

\date{\today }

\maketitle
\flushbottom

\section{Abstract}

We study the stability of a  generalization of the GHS-GM black hole in the presence of a dilaton and a monopole field. We find that the thermal behaviour of system depends on the scalar charge of the dilaton field and as this parameter  is decreasing the system becomes more thermally stable. We also find that, as the charge of the black hole is increasing, both the real and the imaginary parts of the quasi-normal frequencies decrease in absolute value. The   overtone modes die out faster than the fundamental modes and no positive imaginary parts appear, indicating the stabilty of the system.

\tableofcontents

\newpage

\section{Introduction}

If an electromagnetic field is coupled to gravity, very interesting  black hole solutions are generated.  In this  Maxwell-Einstein theory generalized magnetically charged Reissner-Nordstr\"om black hole solutions can be found. In the presence of a dilaton field the electrically charged Garfinkle - Horowitz - Strominger - Gibbons - Maeda (GHS - GM) black hole solutions \cite{Gibbons:1987ps,Garfinkle:1990qj}  are well known. The magnetically charged black holes have been studied and their stability has been investigated; in a magnetically charged Reissner-Nordstr\"om solution it was shown in \cite{Lee:1991qs} that a classical instability is generated by the presence of a magnetic monopole.  A magnetic monopole is a hypothetical particle predicted in string theories \cite{Wen:1985qj}, the existence of which is  questionable, however Dirac has shown that the existence of a magnetic monopole in the Universe implies the quantization of the electric charge \cite{Dirac:1931kp}.

The Euler-Heisenberg theory \cite{Heisenberg:1936nmg} is a generalization of Maxwell theory motivated by the Dirac’s theory in which an electromagnetic field tends to create pairs of particles which leads to a change of Maxwell
equations in the vacuum. Then, as shown in \cite{Obukhov:2002xa}, the magnetization is determined by the clouds of virtual charges surrounding the real currents and charges. Some black holes in this context have been studied in \cite{Karakasis:2022xzm,  Bakopoulos:2024hah, Theodosopoulos:2023ice}.  In these theories the motion of particles in an Euler-Heisenberg, magnetically charged, black hole with a scalar hair has been studied. It was found that the magnetic charge affects the particle motion in a repulsive way. The scalar charge affects the motion in the same manner, while the modified electrodynamics Euler-Heisenberg parameter has minimal effects on the particle motion. A study of the thermodynamics of these black holes was performed  in \cite{Magos:2020ykt, Dai:2022mko}, while the stability of these black holes, calculating the  quasinormal modes, has been studied  in \cite{Breton:2021mju}.

A gravity model was studied in \cite{Kyriakopoulos:2006vy} in the presence of a dilaton and a monopole field. An exact black hole solution was found having  three free parameters. The dominant and the strong energy condition are satisfied outside and on the external horizon the black hole is dressed with scalar hair and the GHS - GM solution is obtained  for certain values of the parameters. The model has been generalized in \cite{Kyriakopoulosb} and an extended model has been constructed \cite{Kyriakopoulosc}, which  also includes electric charges. These models represent generalizations of the (GHS - GM) black hole solutions \cite{Garfinkle:1990qj, Gibbons:1987ps}, they are asymptotically free, and contain more  parameters, which may be tuned to yield alternative theories.

In this work we study the stability of a hairy black hole in the presence of a monopole. We work with a generalization of the model discussed in  \cite{Kyriakopoulosb} and we study its thermal stability \cite{Chatzifotis:2023ioc} and its Quasinormal Modes (QNMs) \cite{Becar:2019hwk}. We do not study the case which involves electric charges, presented in \cite{Kyriakopoulosc}. The hairy black hole we considered has two horizons, an external horizon $r_+$ and an internal one, $r_-,$ where $r_+\ge r_-.$  There is a parameter $c$ that appears in the metric and and a parameter $\ a$ which characterizes the presence of the scalar field, the scalar charge. The decisive parameter for the behaviour of the temperature is this  parameter $\ a$. As the value of the parameter $a$ is decreasing, the system becomes thermally more stable. Also, the calculation of the heat capacity supports this behaviour. Calculating the QNFs we found that their real parts decrease, as $\a$ increases from negative to positive values. The imaginary parts get absolutely smaller, so that the black holes are predicted to live longer. Calculating also the overtone modes we found no positive imaginary parts, so that no instability is generated.  

The work is organised as follows. In Section \ref{theory} we present the basic ingredients of the model we are working on. In Section \ref{thermo} we discuss the thermodynamics of the molel while in  Section \ref{QNMs} we calculate the QNMs. Finaly in \ref{conclusion} we conclude.

\section{Set up of the Theory}
\label{theory}

We consider a gravity theory in the presence of a scalar field with kinetic energy and its coupling to an electromagnetic field. The model we consider is a generalization of the model presented in \cite{Kyriakopoulos:2006vy}. The action is given by
\be S = \int d^4 x \sqrt{-g} L =\int d^4 x \sqrt{-g} \left[R-\f{1}{2} \p_\mu \psi \p^\mu \psi -\left(g_1 e^{(c+\sqrt{c^2+1})\psi} + g_2 e^{(c-\sqrt{c^2+1})\psi}\right) F_{\mu\nu} F^{\mu\nu}\right],\label{ac}
\label{action}\ee 
where 
\be F=Q \sin\theta d\theta \wedge d\phi.\ee is the monopole background.

The solution of the Klein-Gordon, Einstein and Maxwell equations yield the metric
 \be ds^2 = -\f{(r-r_+)(r-r_-)}{r ( r+a)} \left(\f{r}{r+a}\right)^{\f{c}{\sqrt{c^2+1}}} dt^2 + \f{r (r+a)}{(r-r_+)(r-r_-)} \left(\f{r+a}{r}\right)^{\f{c}{\sqrt{c^2+1}}} dr^2 \ee \be + r (r+a) \left(\f{r+a}{r}\right)^{\f{c}{\sqrt{c^2+1}}}(d\theta^2 +\sin^2\theta d\phi^2) \ee 

\be = -f(r) dt^2 + \f{dr^2}{f(r)} + H^2(r) (d\theta^2 +\sin^2\theta d\phi^2),\ee
 where \be f = \f{(r-r_+)(r-r_-)}{r ( r+a)} \left(\f{r}{r+a}\right)^{\f{c}{\sqrt{c^2+1}}},\ H^2 = r (r+a) \left(\f{r+a}{r}\right)^{\f{c}{\sqrt{c^2+1}}},\ee 

while the scalar field profile reads \be e^{\psi} = e^{\psi_0} \left(1+\f{a}{r}\right)^{\f{1}{\sqrt{c^2+1}}},\ee 
where $\psi_0$ is an integration constant. The solution has an external horizon $r_+$ and an internal one, $r_-,$ where $r_+\ge r_-.$  

This is a solution for the system described in (\ref{ac}), provided that
 \be g_1=\f{r_+ r_- (\sqrt{c^2+1}-c)}{2 Q^2 \sqrt{c^2+1}} e^{-(c+\sqrt{c^2+1})\psi_0},\ g_2=\f{(a+r_+)(a+r_-) (\sqrt{c^2+1}+c)}{2 Q^2 \sqrt{c^2+1}} e^{-(c-\sqrt{c^2+1})\psi_0}.
\ee 
The entropy of the black hole is given by \be S = \pi H^2(r_+) = \pi r_+ (r_++a) \left(\f{r_+ + a}{r_+}\right)^{\f{c}{\sqrt{c^2+1}}}.\ee
It is interesting to note that various values of the above parameters can generate known black hole solutions.

(a) If $c=0$ as well as $r_-=0, a=0,$ we get the Schwarzschild black hole solution. The case of $c\ne 0,$ it is a mild generalization of the Schwarzschild.

(b) If $c=0$ as well as $r_+ = r_-,$ then 
$f = \f{(r-r_+)^2}{r^2},\ H^2 = r^2$ and from this metric we get zero temperature and  zero surface gravity.  The inner and outer horizons merge into a degenerate horizon. This is reminiscent, for example, to the extremal Reissner-Nordstrom black holes. In addition, the entropy does not vanish, contrary to the fact  that the entropy should approach zero, if the temperature approaches the absolute zero.

(c) Assume that \be c=0,\ \ a = -r_- = -\f{2 Q^2}{r_+} e^{\psi_0}.\ee 
Then \be g_1=\f{r_+ r_-}{2 Q^2} e^{-\psi_0}\to 1,\ g_2 = \f{(r_+ +a)(r_- + a)}{2 Q^2} e^{\psi_0} \to 0.\ee  
\be 2 M = r_++r_- +a \to r_+,\ee \be e^\psi=e^{\psi_0} \left(1+\f{a}{r}\right) \to e^{\psi_0} \left(1 - \f{ Q^2}{M r} e^{\psi_0}\right)\ee 
\be ds^2 \to - \left(1 - \f{r_+}{r}\right) dt^2 + \left(1 - \f{r_+}{r}\right)^{-1} dr^2 + r \left(r- \f{2Q^2}{r_+}e^{\psi_0}\right) (d\theta^2 +\sin^2\theta d\phi^2) \ee 
\be \to - \left(1 - \f{2 M}{r}\right) dt^2 + \left(1 - \f{2 M}{r}\right)^{-1} dr^2 + r \left(r- \f{Q^2}{M}e^{\psi_0}\right) (d\theta^2 +\sin^2\theta d\phi^2).\ee 
Thus, we get the GHS-GM black hole.

(d) If $c=0,\ a=0,$ we get  $f = \f{(r-r_+)(r-r_-)}{r^2},\ H^2 = r^2,\ \psi = \psi_0.$ 

For generic parameters the scalar curvature reads 
\be R_S=\f{a^2 (r-r_+) (r-r_-)}{2 (c^2+1) r^3 (r+a)^3} \left(\f{r}{r+a}\right)^{\f{c}{\sqrt{c^2+1}}}\ee 
and the Kretschmann scalar reads 
\be R_{\mu\nu\rho\sigma}R^{\mu\nu\rho\sigma} = \f{P(r,a,r_+,r_-)}{4 (c^2+1)^2 r^6 (r+a)^6} \left(\f{r}{r+a}\right)^{\f{2 c}{\sqrt{c^2+1}}},\ee 
where $P(r,a,r_+,r_-)$ is a polynomial with no singularities. Thus, for $r_+>0,\ r_->0,\ a>0,$ there are just coordinate singularities and only at $r=0$ is there a real singularity. If $a<0,$ there will be a real singularity at $r=-a,$ in addition to the point $r=0.$ We will suppose that $r_+>|a|$ for negative $a,$ so that eventual singularities at $r=|a|$ are located within the horizon $r_+.$

For completeness we report the energy-momentum tensor of the model
\be T_{\mu\nu} = \p_\mu\psi \p_\nu\psi +4 f F_{\mu\rho} F_\nu^{\ \rho} -g_{\mu\nu} \left[\f{1}{2} \p_\rho\psi \p^\rho\psi + f F_{\rho\sigma} F^{\rho \sigma}\right].\ee
 Inserting the solution we find 
\be T_{\mu\nu} = \f{a^2}{(c^2+1) r^2 (r+a)^2}\delta_{\mu r} \delta_{\nu r}\ee \be + \left[\f{2 r_+ r_-}{r^2}\left(1-\f{c}{\sqrt{c^2+1}}\right) + \f{2 (r_++a)(r_-+a)}{(r+a)^2} \left(1 + \f{c}{\sqrt{c^2+1}}\right)\right] (\delta_{\mu\theta} \delta_{\nu\theta} +\sin^2\theta \delta_{\mu\phi} \delta_{\nu\phi})\ee
\be - g_{\mu\nu} \left[ \f{a^2 (r-r_+)(r-r_-)}{2 (c^2+1) r^3 (r+a)^3} + \f{r_+ r_-}{r^3 (r+a)} \left(1-\f{c}{\sqrt{c^2+1}}\right) + \f{(r_+ + a) (r_-+a)}{r (r+a)^3} \left(1+\f{c}{\sqrt{c^2+1}}\right)\right] \left(\f{r}{r+a}\right)^{\f{c}{\sqrt{c^2+1}}}.\ee 
The calculation of the eigenvalues of $T_{\mu\nu}$ shows that it satisfies the dominant, as well as the strong energy condition, outside and on the external horizon.


\section{Thermodynamics}
\label{thermo}

In this section calculating the temperature and the heat capacity we study the thermal behaviour of the systeml. 
The temperature is equal to
 \be T=\f{f^\pr(r_+)}{4 \pi} = \f{r_+-r_-}{4\pi r_+ (r_+ +a)} \left(\f{r_+}{r_+ + a}\right)^{\f{c}{\sqrt{c^2+1}}}.\label{TT}\ee 
The asymptotic behaviour of the metric function determines the parameter $M=\f{1}{2} (r_+ + r_- + a).$
Then  the heat capacity of the black hole is given by 
\be C= \f{dM}{dT} = \f{\f{d M}{d r_+}}{\f{d T}{d r_+}} = {\left(\f{r_+ + a}{r_+}\right)^{\f{c}{\sqrt{c^2+1}}} \f{2 \pi \sqrt{c^2+1} r_+^2 (r_+ + a)^2}{\sqrt{c^2+1} (2 r_- - r_+) r_+ + a [(\sqrt{c^2+1} -c) r_- + c r_+]}}.\label{CC}\ee

The heat capacity may become negative; for simplicity we examine the model, in which $c=0.$  In this case the roots of the denominator in $C,$ the roots of equation 
$(2 r_--r_+) r_++a r_-=0$ w.r.t. $r_+$ read $r_- \pm \sqrt{r_- (r_-+a)}.$ 
Since we consider $r_+$ values larger than $r_-,$ so we need only consider $r_- + \sqrt{r_- (r_-+a)}.$ The heat capacity is negative, so the black hole is unstable whenever $r_+$ is greater than $r_- + \sqrt{r_- (r_-+a)},\ \ a\ge r_-.$ Notice that $a,$ if negative, may not be smaller than $-r_-,$ so that the value $a=-r_-$ is an extremal case. 

There are values of $r_+,$ the ones bigger than the peaks, where the temperature has negative slopes. Thus, stability dictates $r_+ <r_-+\sqrt{r_- (r_-+a)},$ equivalently $r_- > \f{r_+^2}{2 r_+ +a}.$ It is to be expected that stability considerations dictate that $r_+$ is smaller than some value, depending on $r_-$ and $a.$ This means that, given $r_+,$ one is not free to vary $r_-$ at will; it should be reasonably close to $r_+.$

\begin{figure}[ht]
\begin{center}
\includegraphics[scale=0.6,angle=0]{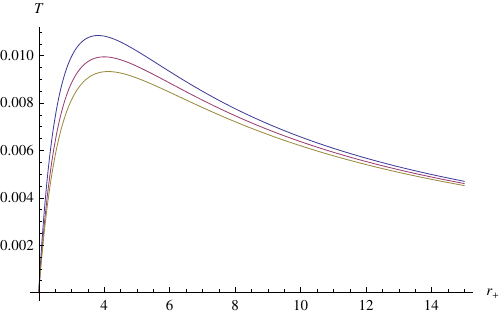}
\includegraphics[scale=0.6,angle=0]{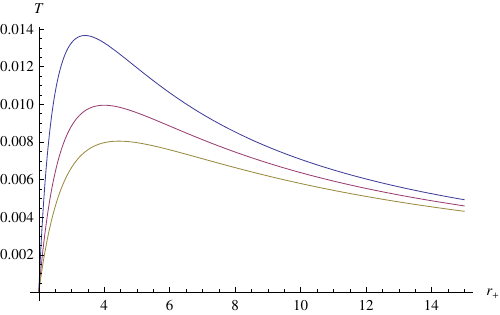}
\includegraphics[scale=0.6,angle=0]{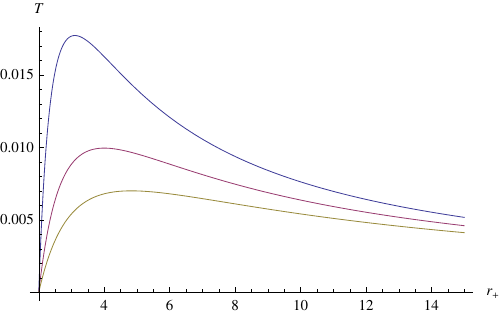}
\end{center}
\caption {Left panels: Temperature versus $r_+$ for $r_-=2$ and $c=-1$ (left panel) $c=0$ (middle panel), and $c=+1$ (right panel). Within each panel one may see the results for $a=-1$ (lowest curve) $a=0$ (middle curve), and $a=+1$ (highest curve).} \label{temp}
\end{figure}

Figure \ref{temp} depicts the temperature versus $r_+$ for $r_-=2.$ The panel on the left represents $c=-1,$ the middle one represents $c=0$ and the right panel corresponds to $c=+1.$ Within each panel the highest curves are the ones with $a-+1;$ then follow the ones for $a=0$ and $a=-1.$  According to the figure the three values of $c$ examined yield similar curves, with just quantitative differences. It does not seem worth insisting on non-zero values for $c$ any more, so in the sequel we concentrate on the case $c=0.$

\begin{figure}[ht]
\begin{center}
\includegraphics[scale=0.6,angle=0]{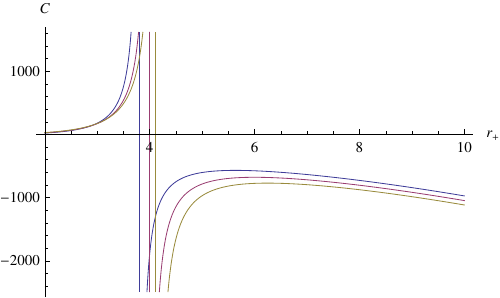}
\includegraphics[scale=0.6,angle=0]{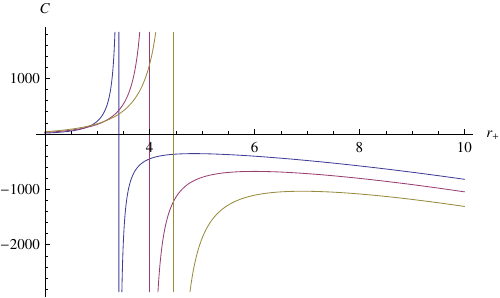}
\includegraphics[scale=0.6,angle=0]{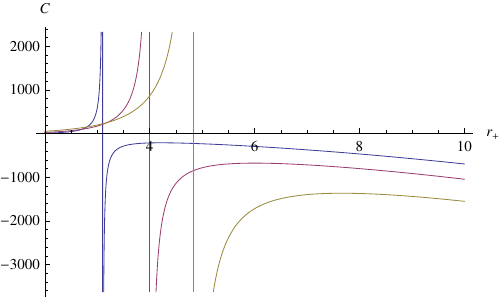}
\end{center}
\caption {Heat capacity versus $r_+$ for $r_-=2$ and $c=-1$ (left panel) $c=0$ (middle panel), and $c=+1$ (right panel). Within each panel one may see the results for $a=-1$ (lowest curve) $a=0$ (middle curve), and $a=+1$ (highest curve).} \label{heatcap}
\end{figure}

The behaviour of the temperature is reflected to the sign of the heat capacity: \be C=\f{d M}{d T}=\f{\f{d M}{d r_+}}{\f{d T}{d r_+}},\ee as depicted in figure \ref{heatcap}. We recall that $M=\f{1}{2}(r_++r_-+a)\Rightarrow \f{d M}{d r_+} = \f{1}{2}.$ The system has negative heat capacity for large enough $r_+,$ which signals thermodynamic instability of "large" black holes. It is interesting that, for $r_-=0,$ the heat capacity is negative for all values of $r_+,$ which is consistent with the well-known behaviour of the Schwarzschild black hole. This novel black hole is similar to the Schwarzschild solution, but it offers a way out from this instability, offering an interval of horizon values, with no instability. However, this comes with a price: $r_+$ may not be too different from $r_-,$ otherwise the instability shows up. 

\begin{figure}[ht]
\begin{center}
\includegraphics[scale=0.8,angle=0]{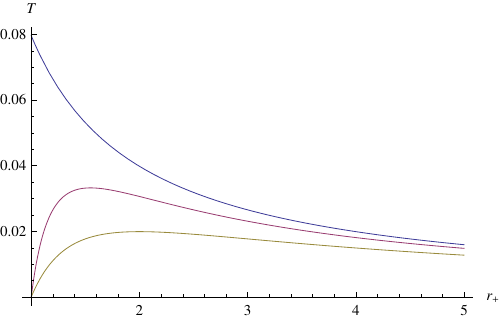}
\includegraphics[scale=0.8,angle=0]{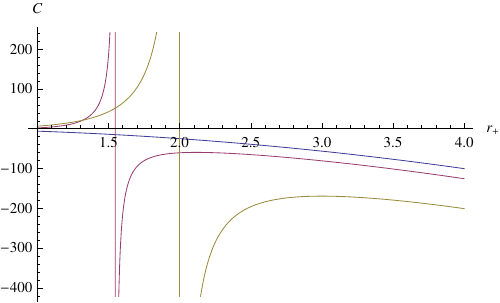}
\end{center}
\caption {Left panel: Temperature  versus $r_+$ for $c=0, r_-=1$ and $a=-1.0$ (highest), $a=-0.7$ and  $a=0.0.$ Right panel: Heat capacity versus $r_+$ for $c=0, r_-=1$ and $a=-1.0$ (highest), $a=-0.7$ and  $a=0.0.$} \label{rm1}
\end{figure}

An interesting special case is depicted in figure \ref{rm1}, left panel, where we depict the temperature versus $r_+$ for $c=0, r_-=1$ and $a=-r_-=-1.0$ (highest), $a=-0.7$ and  $a=0.0.$ The extreme case $a=-r_-$ yields a temperature starting at $T\to \f{1}{4 \pi},$ since $r_+=1,$ and decreases monotonically after this value. This is to be expected, by inspection of equations (\ref{TT}). In the right panel one may find the corresponding heat capacity, which is negative everywhere, so the black hole is unstable. For comparison we also depict the results for $a=-0.7;$ in the left panel the temperature starts with a positive slope, attains a maximum and then it decreases. The corresponding heat capacity on the right hand panel is positive at first and at some point it has an infinite discontinuity and dives to negative values. Similar behaviour is found in the third case $a=0.$

\section{QNMs}
\label{QNMs}

An important tool for study  the stability of a hairy black hole is to calculate the QNMs. In \cite{Holzhey:1991bx} it was argued that a thermal description is inadequate for extreme holes of the charged dilaton family. The reason is that these black holes can have
zero entropy but non-zero, and even  formally infinite, temperature.
To cure this problem they  analyzed the
perturbations around the extreme holes and they  showed that these holes are protected by  alternatively potential barriers, which remove them from thermal contact with the external world. In \cite{Ferrari:2000ep} the perturbations of the charged, dilaton black hole described by the GHS - GM black hole solutions were studied while in \cite{Lin:2010zzf} the gravitational perturbation of GHS - GM dilaton black hole and QNMs were calculated. Calculations of the QNFs  have been done through the third order WKB methodology pioneered in
\cite{Iyer:1986np,Iyer:1986nq}

We have already seen that $r_+$ and $r_-$ may not be too far from each other, due to thermodynamic stability considerations. However, variation of the inner horizon $r_-$ or the $a$ parameter influences the black hole properties. We have chosen to fix $c$ to zero, $r_+$ to $+5,$ and examine the influence of $r_-$ on the QNMs for $a=-\f{r_+}{5},\ 0,\ \f{r_+}{5}.$ For condition $r_->\f{r_+^2}{2 r_+ +a}\Rightarrow \f{r_-}{r_+} > \f{1}{2 + \f{a}{r_+}}$ informs us that, for given $a$ and $r_+,$ the inner horizon should be larger than some limit. If $a=\f{r_+}{5},$ the lowest value of $\f{r_-}{r_+}$ should be $\f{5}{11};$ if $a=0,$ the lowest value should be $\f{1}{2},$ while, if $a=-\f{r_+}{5},$ $r_-$ should be $\f{5}{9}.$ We chose some values for $\f{a}{r_+}$ which are permitted and calculated the corresponding QNFs. 

We now sketch the derivation of the axial scalar perturbation potential. We recall the background
\be f = \f{(r-r_+)(r-r_-)}{r ( r+a)} \left(\f{r}{r+a}\right)^{\f{c}{\sqrt{c^2+1}}},\ H^2 = r (r+a) \left(\f{r+a}{r}\right)^{\f{c}{\sqrt{c^2+1}}},\ \ e^{\psi} = e^{\psi_0}\left(1+\f{a}{r}\right)^{\f{1}{\sqrt{c^2+1}}}.\ee  
The axial perturbations take the form 
\be h_{\mu\nu}^{axial} = \left(\begin{array}{cccc} 0 & 0 & 0 & h_0(r) \\  0 & 0 & 0 & h_1(r)  \\ 0 & 0 & 0 & 0 \\ h_0(r) & h_1(r) & 0 &0 \\ \end{array}\right) \sin\theta \f{\p Y_{lm}}{\p\theta} e^{-i \omega t}.\ee  We next consider equations \be \delta G_{t \theta}=0,\ \delta G_{t \phi}=0,\ \delta G_{r \theta}=0,\ \delta G_{r \phi}=0,\ee 
which yield equations involving $h_1(r)$ and $h_0(r),$ which may be eliminated through these relations. 
Defining the variable \be \Psi =\f{f(r)}{H(r)} h_1(r),\ee yields the equation
 \be \f{d^2 \Psi}{d r_*^2} +(\omega^2-V_{axial}(r)) \Psi=0,\ \ V_{axial}(r) = f(r) \left[\f{l (l+1)}{H^2}-\f{3 f^\pr(r) H^\pr(r)}{H(r)} \right],\ \ dr_*=\f{d r}{f(r)}.\label{VV}\ee 
We will extensively use the variable \be Q(r) \equiv \omega^2-V_{axial}(r),\ee which yield
 \be \f{d^2 \Psi}{d r_*^2} +Q(r) \Psi=0.\ee 



For axial modes only the gravitational degrees of freedom fluctuate, because the non-zero component of the background field strength, namely $F_{\theta\phi},$ is proportional to $\sin\theta,$ which is parity even, so it couples only to parity even perturbations. Dilaton does not couple to axial perturbations either for the same reasons, since it is spherically symmetric. Thus one ends up with an equation which involves only gravitational degrees of freedom. 


In our generalized case
 \be H(r) = \sqrt{r (r+a)} \left(\f{r+a}{r}\right)^{\f{c}{2 \sqrt{c^2+1}}}\Rightarrow \f{dH}{dr} = \f{\left(\f{r+a}{r}\right)^{\f{c}{2 \sqrt{c^2+1}}} [a (\sqrt{c^2+1} - c) + 2 \sqrt{c^2+1} r]}{2 \sqrt{c^2+1} \sqrt{r (r+a)}}.\ee
 Notice that, since we are basically dealing with gravitational perturbations, $l$ should be greater or equal to 2. 

To make a consistency check we  set $a\to 0,\ c\to 0, \ r_-\to 0$ and then one get
: $H\to r,\ H^\pr \to 1,\ f(r)\to 1-\f{r_+}{r},$ 
so that
 \be V_{axial} \to f(r) \left[\f{l (l+1)}{r^2} - \f{3 r_+}{r^3}\right].\ee Setting $r_+\to 2 M,$ we get the well known result of the effective potential of axial perturbations
 \be V_{axial} = f(r) \left[\f{l (l+1)}{r^2} - \f{6 M}{r^3}\right].\ee

Following \cite{Iyer:1986np,Iyer:1986nq}, one has to spot the point $r_*=r_{* 0}$ of the maximum, of $-Q(r_*):$
\be \left. -\f{d Q(r_*)}{d r_*}\right|_{r_{* 0}} = 0.\ee
Define the quantities 
\be \Lambda(n) = -\f{i}{(2 Q_0^\pp)^{1/2}} \left[\f{1}{8} \f{Q_0^{(4)}}{Q_0^\pp} \left(\f{1}{4}+\alpha^2\right)-\f{1}{288} \left(\f{Q_0^{(3)}}{Q_0^\pp}\right)^2 (7+60 \alpha^2)\right],\ \ \alpha\equiv n+\f{1}{2},\ee
\be \Omega(n) =\f{1}{2 Q_0^\pp} \left[\f{5}{6912} \left(\f{Q_0^{(3)}}{Q_0^\pp}\right)^4 \left(77+188 \alpha^2\right)-\f{1}{384} \left(\f{(Q_0^{(3)})^2 Q_0^{(4)}}{(Q_0^\pp)^3}\right) (51+100 \alpha^2)\right.\ee
\be \left. +\f{1}{2304} \left(\f{Q_0^{(4)}}{Q_0^\pp}\right)^2 (67+68 \alpha^2)+\f{1}{288} \left(\f{Q_0^{(3)} Q_0^{(5)}}{(Q_0^\pp)^2}\right)(19+28\alpha^2) - \f{1}{288} \left(\f{Q_0^{(6)}}{Q_0^\pp}\right)(5+4\alpha^2) \right].\ee 
where  \be Q_0^{(i)} = \left. \f{d^i Q}{d r_*^i}\right|_{r_*=r_{*0}}.\ee
Next we calculate the square root of \be \omega^2 = [V_0+(-2 V_0^\pp)^{1/2} \Lambda(n)]-i \left(n+\f{1}{2}\right) (-2 V_0^\pp)^{1/2} [1+\Omega(n)].\ee 

We need to calculate derivatives with respect to $r_*.$ We may either express $Q$ as a function of $r_*,$ or express the derivatives with respect to $r_*$ in terms of derivatives with respect to $r.$ For example
\be \f{d Q}{d r_*} = \f{d Q}{d r} \f{d r}{d r_*} = f \f{d Q}{d r},\ \f{d^2 Q}{d r_*^2} = f \f{d}{d r} \left(f \f{d Q}{d r}\right).\ee 

\begin{figure}[ht]
\begin{center}
\includegraphics[scale=0.6,angle=0]{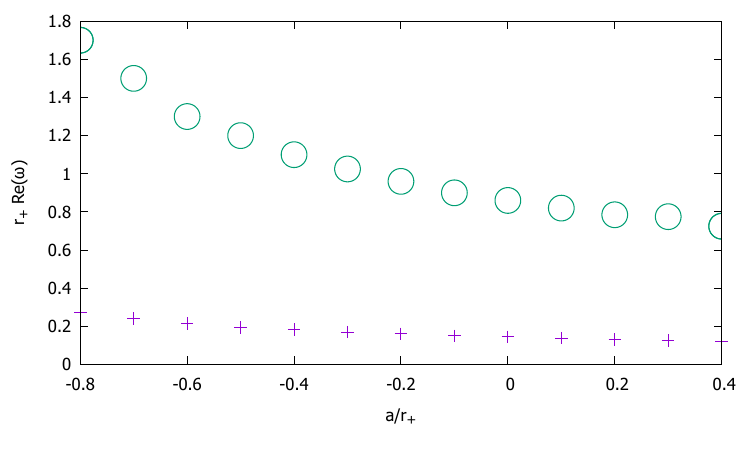}
\includegraphics[scale=0.6,angle=0]{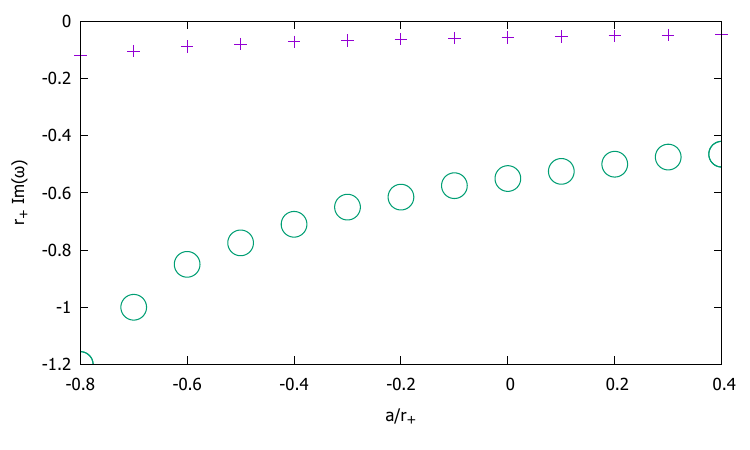}
\end{center}
\caption {Quasinormal frequencies $r_+ \omega$ versus the $\f{a}{r_+}$ parameter for $c=0,\ r_+=5,\ l=2.$ Left panel: the real part $r_+ \Re(\omega).$ Right panel: the imaginary part $r_+ \Im(\omega)$. Results for fundamental (denoted by crosses) and first overtone (denoted by circles) states are displayed.} \label{l2}
\end{figure}

Finally the QNFs are given by the formula
\be r_+ \omega = r_+ \sqrt{ [V_0+\sqrt{-2 V_0^\pp} \tilde{\Lambda}(n)] - i \left(n+\f{1}{2}\right) \sqrt{-2 V_0^\pp}  [1+\tilde{\Omega}(n)]}.\ee
In figure \ref{l2} we depict the quasinormal frequencies $r_+ \omega$ versus $a$ for a black hole for which $c=0,\ l=2, \ r_+=5,\ \f{r_-}{r_+} =\f{1}{2 +\f{a}{r_+}}.$ That is we used the smallest value allowed for $\f{r_-}{r_+}.$ On the left panel the real parts $r_+\Re(\omega)$ is depicted for the fundamental (lower points, denoted by crosses) and the first overtone (upper points, denoted by circles). The imaginary parts $r_+ \Im(\omega)$ are depicted in the right panel: results for the fundamental are represented by crosses and those for the overtone are given by circles. We find out that the quantities $r_+ \Re(\omega)$ decrease algebraically, as $\f{a}{r_+}$ increases from negative to positive values; on the contrary, the imaginary parts $r_+ \Im(\omega)$ decrease absolutely for increasing $\f{a}{r_+},$ signaling longer - lived states. The overtone values are more negative than the values of the fundamental modes. Both become less negative for increasing $\f{a}{r_+}.$

\begin{figure}[ht]
\begin{center}
\includegraphics[scale=0.6,angle=0]{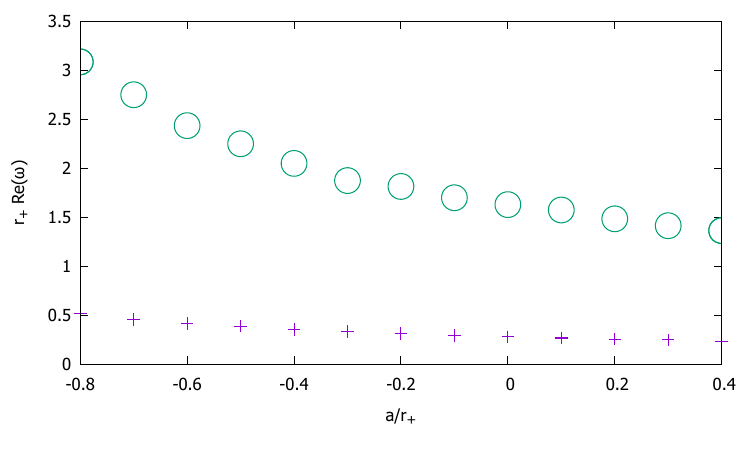}
\includegraphics[scale=0.6,angle=0]{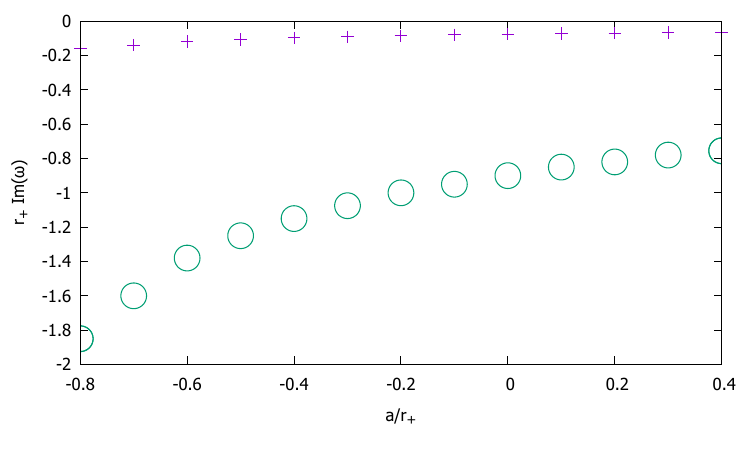}
\end{center}
\caption {Quasinormal frequencies $r_+ \omega$ versus the $a$ parameter for $c=0,\ r_+=5,\ l=4.$ Left panel: the real part $r_+ \Re(\omega).$ Right panel: the imaginary part $r_+ \Im(\omega)$. Results for fundamental (denoted by crosses) and first overtone (denoted by circles) states are displayed.} \label{l4}
\end{figure}

Similar behaviour is observed for $l=4,$ which is depicted in figure \ref{l4}. In both cases the overtone modes die out faster than the fundamental modes. No positive imaginary parts appear, so that no instability of this sort appear in the cases we have been studying.

\section{Conclusions}
\label{conclusion}

We have studied a black hole solution with two horizons in the presence of a monopole and a dilaton field. This solution is a generalization of the GHS - GM solution. We have found that the model, if the internal horizon $r_-$ and the scalar charge $a$ are given is stable for small enough external horizon $r_+$ and unstable for larger values.  We also calculated the QNFs for axial perturbations (taking care that we remain in the stable region) and found that the real parts of the QNFs decrease for increasing scalar charge  that is $a$ causes the oscillations to be slower. The imaginary parts of the QNFs algebraically increase for larger scalar charge $a,$ that is they are longer lived; in addition they are always negative, so no instability of this kind is found.

We have also calculated some overtones (we depicted the first ones). The real part of the of QNFs are bigger, i.e. the oscillations are faster than the corresponding ones of the fundamental states, while the imaginary part of  quantities  of QNFs for the first overtone are absolutely larger than the fundamental states: the overtones die out faster than the fundamental states. These results have been found for angular numbers $l=2$ and $l=4.$ In the latter case the first overtone gets absolutely larger values than the fundamental, so the $l=4$ states die out faster than the $l=2$ states.

\end{document}